%% file: main.tex
\documentclass{article}

\usepackage{arxiv}
\usepackage[utf8]{inputenc} 
\usepackage{url} 
\usepackage{tikz} 
\usepackage{amsmath} 
\usepackage[capposition=top]{floatrow} 
\usepackage[flushleft]{threeparttable} 
\usepackage{booktabs} 
\usepackage{dcolumn} 
\usepackage[labelfont=bf]{caption} 
\usepackage{authblk} 
\usepackage{hyperref}

\title{Adjusting for misclassification of an exposure in an individual participant data meta-analysis}

\author[1,2,3]{Valentijn M.T. de Jong}

\author[4]{Harlan Campbell}

\author[5]{Lauren Maxwell}

\author[5,6,7]{Thomas Jaenisch}

\author[4]{Paul Gustafson}

\author[1,2]{Thomas P.A. Debray}

\affil[1]{\footnotesize Julius Center for Health Sciences and Primary Care, University Medical Center Utrecht, Utrecht University, Utrecht, the Netherlands}
\affil[2]{\footnotesize Cochrane Netherlands, University Medical Center Utrecht, Utrecht University, Utrecht, the Netherlands}
\affil[3]{\footnotesize Data Analytics and Methods Task Force, European Medicines Agency, Amsterdam, the Netherlands}
\affil[4]{\footnotesize Department of Statistics, University of British Columbia, Vancouver, British Columbia, Canada}
\affil[5]{\footnotesize Heidelberg Institute of Global Health, Heidelberg Medical School, Heidelberg University, Heidelberg, Germany}
\affil[6]{\footnotesize Center for Global Health, Colorado
School of Public Health, Aurora, Colorado, United States of America,}
\affil[7]{\footnotesize Department of Epidemiology,
Colorado School of Public Health, Aurora, Colorado, United States of America}

\date{\today}

\begin{document}

\maketitle
\begin{abstract}
A common problem in the analysis of multiple data sources, including individual participant data meta-analysis (IPD-MA), is the misclassification of binary variables. 
Misclassification may lead to biased estimates of model parameters, even when the misclassification is entirely random. We aimed to develop statistical methods that facilitate unbiased estimation of adjusted and unadjusted exposure-outcome associations and between-study heterogeneity in IPD-MA, where the extent and nature of exposure misclassification may vary across studies. 

We present Bayesian methods that allow misclassification of binary exposure variables to depend on study- and participant-level characteristics. In an example of the differential diagnosis of dengue using two variables, where the gold standard measurement for the exposure variable was unavailable for some studies which only measured a surrogate prone to misclassification, our methods yielded more accurate estimates than analyses naive with regard to misclassification or based on gold standard measurements alone. In a simulation study, the evaluated misclassification model yielded valid estimates of the exposure-outcome association, and was more accurate than analyses restricted to gold standard measurements. 

Our proposed framework can appropriately account for the presence of binary exposure misclassification in IPD-MA. It requires that some studies supply IPD for the surrogate and gold standard exposure and misclassification is exchangeable across studies conditional on observed covariates (and outcome). 
The proposed methods are most beneficial when few large studies that measured the gold standard are available, and when misclassification is frequent. 
\end{abstract}


\input{text}

\bibliography{references}
\bibliographystyle{vancouver}

\newpage
\input{Ch_me/denv/appendix_denv_ofi}

\end{document}

%% file: text.tex
\section{Introduction}
Individual participant data meta-analysis (IPD-MA) comprises the pooling and subsequent analysis of the participant-level data from multiple studies. As an IPD-MA summarizes the evidence through synthesis and analysis of all data available to answer a specific research question, it is generally seen as the highest standard of scientific evidence  \cite{stewart_ipd_2002}.  It is therefore unsurprising that IPD-MA have become increasingly common to summarize the evidence from experimental and observational studies, and that their results can substantially impact clinical practice. Although IPD-MA are frequently conducted to study the efficacy of therapeutic interventions, they can also be used to investigate etiologic, diagnostic, and prognostic variables. In observational research, data are commonly gathered using methods or instruments that are prone to measurement error (ME), but this may also occur in randomized controlled trials (RCTs). \cite{sjoding_acute_2016,
simmons_assessment_2015,
choudhry_randomized_2017}

Measurement error is any difference between the value that is observed for a variable and its true value. Measurement error may arise due to a variety of random or systematic causes, such as errors in measurement instruments or their application, the reading of such instruments, poor recall, misunderstanding items on questionnaires and data entry and management. The presence of measurement error may introduce (upward or downward) bias in estimates of parameters, even when the error is entirely random and independent of other variables. \cite{keys_effect_1963, carroll_measurement_2006, gustafson_misclassification_2014}  

Measurement error in categorical variables is referred to as misclassification. It is commonly believed that misclassification of the exposure leads to attenuation of exposure-outcome associations. \cite{van_smeden_five_2019} As a result, researchers often interpret estimates as conservative and dismiss the need for more advanced analyses that account for ME. \cite{brakenhoff_measurement_2018} However, attenuation is only guaranteed to occur when the misclassification is non-differential (that is, misclassification is independent of the outcome given the measured covariates),  \cite{bross_misclassification_1954, keys_effect_1963, copeland_bias_1977, gustafson_measurement_2003, gustafson_misclassification_2014} the exposure has no more than two categories \cite{dosemeci_does_1990, birkett_effect_1992} and all covariates are measured without error. \cite{carroll_measurement_2006} When a covariate is also measured with error, the bias introduced by including the mismeasured covariate in a multivariable regression analysis becomes much more difficult to quantify. \cite{carroll_measurement_2006} Further, extreme misclassification can reverse the sign of the observed association. \cite{weinberg_when_1994}

In an individual participant data meta-analysis (IPD-MA), misclassification may be present in one or more studies. For instance, when the IPD from previously published studies are combined, a less accurate measurement instrument for a certain exposure variable may have been used in some studies. If one of these instruments is prone to misclassification, this will result in a biased  estimate for the corresponding exposure's effect. Therefore, in IPD-MA it is generally recommended to standardize measurements, and where possible to adjust for misclassification to reduce bias. \cite{debray_evidence_2019, campbell_measurement_2021}

In meta-analysis, methods must also account for the effects of clustering in individual studies \cite{abo-zaid_individual_2013} and should allow for heterogeneity of the effect of interest. Hence, methods that account for misclassification must do so as well. Further, it may occur that different measurements methods are used across studies. This directly implies that a gold standard measurement may be missing for entire studies. Applying a traditional method that accounts for misclassification in IPD-MA therefore requires that the misclassification rate is transportable to other studies. This may be tenable when the measurement instruments, protocol, population and setting are the same in the included studies, but this would be a rare occasion in the context of IPD-MA. Hence, a method that accounts for possible heterogeneity across studies in misclassification as well as outcome prevalence and the exposure-outcome association should then be applied.


In this article, we consider a binary exposure in an IPD-MA that is prone to misclassification error. We distinguish between measurements that are obtained (or defined) according to the gold standard, and measurements that are made using an instrument that is prone to error (further referred to as the surrogate exposure). We subsequently discuss how valid inferences (at least to a certain degree) can be made while the gold standard measurements for the exposure are missing in some studies, using information on the surrogate exposure and the observed participant characteristics. We adopt a Bayesian estimation framework that extends previously proposed methods \cite{falley_bayesian_2018, nelson_bayesian_2018, lian_bayesian_2019} for addressing misclassification in single studies and in aggregate data meta-analysis (AD-MA).

In section \ref{sec:example_denv:intro} we provide our motivating example of the diagnosis of the dengue virus. In section \ref{sec:misclass_methods} we discuss existing methods for dealing with misclassification, and provide our extensions thereof. We apply these methods in section \ref{sec:example_denv:application} and provide a discussion in section \ref{sec:discuss_misclass}.
 
\section{Motivating example: Diagnosing dengue}
\label{sec:example_denv:intro}
An estimated 100 million infections of dengue occur globally each year. \cite{jaenisch_clinical_2016}  Although dengue infection is often asymptomatic, it can also be fatal and patients can present with various clinical symptoms ranging from mild febrile illness to hemorrhagic fever, organ impairment and hypovolaemic shock. \cite{anders_epidemiological_2011,jaenisch_clinical_2016} In its early phase, dengue can be difficult to distinguish from other febrile illnesses (OFI) such as influenza, chikungunya, measles, leptospirosis and typhoid due to the similarity of clinical symptoms, which include headache and rash. Therefore, the identification of laboratory and other clinical variables that aid in the differential diagnosis of dengue is imperative. \cite{jaenisch_clinical_2016} In this motivating example we focus on the strength of the association between muscle pain and dengue vs OFI, conditional on the presence of joint pain. 

To assess the added diagnostic value of muscle pain in the differential diagnosis of dengue vs OFI, a multivariable logistic prediction model can be developed.  Suppose several studies have fit such models to data and that for some studies the presence of muscle pain data has been tainted by misclassification.  In order to show the potential impact of the misclassification, we

 use simulated IPD for 10 studies (Figure \ref{sim:fig:dengue_bar_chart}), that are based on real data gathered in three cross-sectional studies of the IDAMS consortium (see Supporting Information) that aimed to improve the differential diagnosis of dengue. \cite{jaenisch_clinical_2016} The IPD were generated according to three scenarios with varying heterogeneity in the outcome model. 

In the first scenario we defined the heterogeneity parameters such that all studies have the same (true) prevalence of dengue conditional on the exposure and covariate and the same (true) exposure-outcome association of muscle pain, conditional on the covariate joint pain. In the second scenario we allowed for heterogeneity in the true prevalence of dengue conditional on the exposure and covariate but not in the true exposure-outcome association, conditional on the covariate. In the third scenario we allowed for the presence of heterogeneity in both the true prevalence of dengue conditional on the exposure and covariate as well as the true exposure-outcome association of muscle pain, conditional on the covariate. This third scenario resembles the real world scenario where the diagnosis of dengue is made more difficult by the possible presence of Chikungunya, which is also a febrile illness. 

As Chikungunya is associated with (often even worse) muscle pain, \cite{sudeep_chikungunya:_2008} the association between muscle pain and dengue may be smaller (or entirely absent) in studies where Chikungunya is prevalent, compared to studies where it is not present. In all scenarios we allowed the true prevalence of muscle pain and the true misclassification rates to vary across studies. The challenge is to account for this rate of misclassification that is heterogeneous across studies and depends on patient covariates, while simultaneously accounting for heterogeneity in the prevalence of dengue, and heterogeneity in the muscle pain-dengue association. In the following sections we first provide a short overview of methods for accounting for misclassification in single studies and in AD-MA before we move on to accounting for these sources of heterogeneity in IPD, such in this IPD-MA of the muscle pain-dengue association.

\input{Ch_me/denv/dengue_plot}

\section{Methods}
\label{sec:misclass_methods}
Many methods have been developed to adjust for misclassification of exposures in  the analysis of a single study.  These include regression calibration and multiple imputation based methods.  Methods for adjusting meta-analyses of aggregate data for misclassification have also been proposed. We start by briefly summarizing these methods and their characteristics. More detailed information is available from Keogh et al. \cite{keogh_toolkit_2014}

\subsection{Adjusting for misclassification in single studies}

\subsubsection{Regression calibration}
\label{sec:methods;sub:reg_cal}
In regression calibration, the outcome is regressed on the expected value of the exposure, given the surrogate exposure and covariates. The expected value of the exposure can be estimated by regressing the exposure on the surrogate exposure and covariates for participants for whom all these variables have been measured. When modeling a continuous outcome with linear regression this approach may yield unbiased estimates of the exposure-outcome association. \cite{carroll_measurement_2006} However, regression calibration has been demonstrated to yield (somewhat) biased results when applied to logistic regression  \cite{carroll_measurement_2006, cole_multiple-imputation_2006, keogh_toolkit_2014}. 
As regression calibration does not use the observed outcome for estimating the expected value of the exposure, it cannot account for differential misclassification.

\subsubsection{Multiple imputation}
\label{sec:methods;sub:mi}
In the multiple imputation approach, gold standard and surrogate measurements of the exposure are treated as separate covariates. Participants for whom the gold standard or surrogate measurement has not been applied are then considered to have missing values for the corresponding covariate(s). If there are sufficient participants for whom the surrogate and gold standard exposure are available, the missing exposures can be imputed.

Multiple imputation for measurement error (MIME) is an implementation of multiple imputation (MI), which was designed to deal with missing data. In MI using chained equations, each variable (or a transformation thereof) is iteratively regressed on all other variables. The estimated regression models are then used to impute missing values. In MIME, the estimated regression models are used to impute the missing gold standard measurement of the exposure for participants for whom only the surrogate is observed.  MIME models typically include the outcome as covariate, which naturally accounts for differential error if the imputation model is correctly specified. However, it overestimates the uncertainty in the imputation of the true exposure. \cite{cole_multiple-imputation_2006}

\subsection{Adjustment for misclassification in a meta-analysis of contingency tables}
Most meta-analyses are based on aggregate data. When the exposures are binary, the aggregate data for the exposure-outcome associations are often presented as counts in contingency tables. Provided that contingency tables for the surrogate-gold standard exposure association are also available, one can adjust for the misclassificication in the surrogate exposure-outcome association that is unadjusted for covariates. \cite{lian_bayesian_2019} 

\subsubsection{Exchangeability in meta-analysis}
As the rate of misclassification may differ across studies, Lian et al recently developed a model that accounts for clustering and heterogeneity and relaxed the assumption of transportability to exchangeability. \cite{lian_bayesian_2019} That is, the degree of misclassification is allowed to vary across studies by applying a random effect. The resulting coefficients for the misclassification model and for the exposure-outcome model need not come from the same studies if exchangeability can be assumed. This is advantageous, as it implies that studies in which misclassification was not investigated can be included in the analysis.

Although the model of Lian et al does not assume that misclassification in the measured exposure is common across studies, the exchangeability assumption nevertheless requires that misclassification is independent of any patient-level covariates, given the value of the gold standard measurement of the exposure. \cite{lian_bayesian_2019} In particular, the exchangeability assumption implies that misclassification is assumed to depend solely on study-level variables. This is an important distinction, as misclassification that is non-differential given covariates, may be differential when these covariates are not taken into account. \cite{gustafson_misclassification_2014} Thus, if misclassification rates are different for the levels of the outcome and patient-level covariates can explain those differences, then these covariates must be taken into account.

\subsection{Adjustment for misclassification in AD-MA}
Extending methods that rely on stratified contingency tables to the analysis of covariate-adjusted exposure-outcome associations may be impractical. It would require that studies provide contingency tables that are stratified for the outcome, gold standard measurement of the exposure, surrogate exposure and every adjustment variable. Clearly, this may be infeasible for a large number of variables. 
Alternatively, one may opt to adjust for misclassification in a meta-analysis of aggregate data, that is, using exposure-outcome associations (and standard errors) reported in the form of regression coefficients such as (log) risk or odds ratios that have been adjusted for covariates. If all of these reported estimates (including the standard errors) are appropriately adjusted for misclassification in their respective studies, one could analyze these with traditional meta-analysis methods. On the other hand, if the estimation of these covariate adjusted exposure-outcome associations did not include accounting for misclassification, then this would have to occur in the meta-analysis.

If IPD are available for the gold standard and surrogate measurements of the exposure, one might apply a misclassification model to adjust the reported exposure-outcome associations for misclassification, but this would require the assumption of exchangeability of misclassification across the included studies. \cite{lian_bayesian_2019} This assumption would clearly be violated in case the misclassification is dependent on participant-level covariates. For instance, in our motivating example the misclassification of muscle pain was associated with the participant-specific value of joint pain. If the measurement for joint pain is missing for a participant, then the information to estimate the expected value of the missing measurement of muscle pain is missing for that participant. In the case of AD-MA, this implies that the covariate joint pain would be missing for the entire study. Thus, any participant-specific misclassification would not be accounted for. In the next section we describe how the exchangeability assumption in meta-analysis models for misclassification can be relaxed if IPD are available.

\subsection{Adjustment for misclassification in a meta-analysis of individual participant data}
\label{sec:meth;subsec:ipd}
We extend the methods of Nelson et al \cite{nelson_bayesian_2018} and Lian et al \cite{lian_bayesian_2019} to incorporate participant level covariates in a one-stage IPD-MA for potentially misclassified binary exposures. As such, we allow the probability of misclassification to depend on study-level variables and on individual participant level covariates that are observed without error. Further, modeling of IPD allows us to estimate the adjusted (i.e. multivariable) exposure-outcome associations. For example, suppose that misclassification of muscle pain may occur in the differential diagnosis of dengue. 

Let $x_{ij}$ denote the gold standard measurement of the binary exposure (e.g. muscle pain) for participant $i, i = 1,...,n_j$ in study $j$, $j = 1, ...,J$. The surrogate exposure is given as $x^*_{ij}$  and represents a possibly misclassified measurement of the exposure. We assume that $x^*_{ij}$ and $x_{ij}$  have been observed for some participants in some studies, and that for some participants in some studies both have been observed. Further, we assume that $z_{ij}$ is a covariate (e.g. joint pain) without measurement error and that $y_{ij}$ is a binary outcome (e.g. dengue).

Following the approach described by Richardson and Gilks \cite{richardson_bayesian_1993}, we specify three submodels to account for misclassification: a measurement model, an exposure model and an outcome model. In the measurement model, the surrogate exposure (i.e. the measurement of the exposure that is prone to misclassification) is predicted, conditional on the latent gold standard measurement of the exposure, to determine the extent of misclassification. The measurement model models the relation $x^*_{ij} \sim f(x_{ij}, z_{ij})$. In the exposure model, the latent gold standard measurement of the exposure is regressed on covariates that are measured without error, in order to predict the gold standard measurement of the exposure in participants for whom it is missing. Hence, the exposure models the relation $x_{ij} \sim f(z_{ij})$. Note that the exposure model is commonly referred to as the exposure model in etiological studies where the gold standard measurement of the exposure is missing. In the outcome model, the outcome is regressed on the latent gold standard measurement of the exposure and on covariates that are measured without error, to determine the exposure-outcome relationship. The outcome model models the relation $y_{ij} \sim f(x_{ij}, z_{ij})$. Although our model generalizes to multiple covariates, we restrict our notation to a single covariate for simplicity.

\subsubsection{Common effects IPD-MA}
We start with describing an IPD-MA misclassification model containing three submodels that assumes common effects across studies. Hence, all data are analysed as if they were measured in a single study. In this first model, the probability of misclassification only depends on the value of the gold standard measurement of the exposure. The measurement (sub)model is then given by:

\begin{equation}
\label{eq:x*_x}
    \begin{split}
        x^*_{ij} & \sim \textnormal{Bernoulli}(p^*_{ij}),\\
        g(p^*_{ij})& = \lambda x_{ij} + \phi (1 - x_{ij})\\
    \end{split}
\end{equation}

\noindent where $\lambda \sim N(0, \sigma^2_\lambda), \phi \sim N(0, \sigma^2_\phi)$ and $g(.)$ is a link function. For instance, one could choose the logit for $g(.)$, such that intercept parameters represent log odds and (exposure) coefficient parameters represent log odds ratios. This is equivalent to a measurement model proposed by Nelson et al, \cite{nelson_bayesian_2018} as $\lambda$ and $\phi$ are parameters that determine  ${g(\textnormal{sensitivity})}$ and ${g(1 - \textnormal{specificity})}$, respectively. The above parametrization allows us to introduce covariates to the measurement model in subsequent steps. We leave the variance parameters unspecified, as fixed values may be supplied for these, though one may also supply prior distributions for the variance parameters.

The exposure model aims to estimate the relationship between the gold standard measurement of the exposure and covariate(s). It is simultaneously applied to predict the probability that the exposure is present in participants for whom the gold standard measurement of the exposure status is missing. For participants for whom the gold standard measurement of the exposure status is missing, the expected value given covariates is imputed following this model. It is given by:

\begin{equation}
\label{eq:x_z}
    \begin{split}
        x_{ij} & \sim \textnormal{Bernoulli}(p_{ij}),\\
        g(p_{ij}) & = \gamma_0 + \gamma_{1} z_{ij}, \\
    \end{split}
\end{equation}

\noindent where $\gamma_{0} \sim N(0, \sigma^2_{\gamma_0})$ and $\gamma_{1} \sim N(0, \sigma^2_{\gamma_1})$. Thirdly, of course, we describe the model that is designed to assess the (adjusted) exposure-outcome association. This outcome model is given by:
\begin{equation}
\label{eq:y_x+z}
    \begin{split}
        y_{ij} & \sim \textnormal{Bernoulli}(\pi_{ij}),\\
        g(\pi_{ij}) & = \beta_0 + \beta_1 z_{ij} + \beta_2 x_{ij}, \\
    \end{split}
\end{equation}

\noindent where $\beta_{0} \sim N(0, \sigma^2_{\beta_{0}}), \beta_{1} \sim N(0, \sigma^2_{\beta_{1}}), \beta_{2} \sim N(0, \sigma^2_{\beta_{2}}), \beta_0$ is an intercept, $\beta_1$ is the coefficient for the covariate and $\beta_2$ is the coefficient (log odds ratio) for the exposure of interest. Equations \ref{eq:x*_x}, \ref{eq:x_z} and \ref{eq:y_x+z} together make up the least complex misclassification model that we consider here and are illustrated in Figure \ref{fig:model_basic_and_adv}. The likelihood of this model is given by the product of the likelihoods of the three submodels, including their priors: 

\begin{equation}
            p\big(\lambda, \phi\big) p \big(\gamma_0, \gamma_1\big) p \big(\beta_0, \beta_1, \beta_2\big)
    \prod_j \prod_i p \big(x^*_{ij}|x_{ij}, \lambda, \phi\big) 
    \prod_j \prod_i p \big(x_{ij}  |z_{ij},         \gamma_0, \gamma_1\big) 
    \prod_j \prod_i p \big(y_{ij}  |x_{ij}, z_{ij}, \beta_0, \beta_1, \beta_2\big)
\end{equation}

Although the implementation of aforementioned misclassification models is fairly straightforward in an IPD-MA, their justification becomes problematic when studies differ with respect to case-mix, baseline risk, exposure-outcome associations or the extent of misclassification. We therefore discuss how to adjust the submodels accordingly.

\input{figure_models}

\subsubsection{Accounting for between-study heterogeneity in the distribution of the exposure}
A common situation in IPD-MA is the presence of heterogeneity in case-mix distributions. \cite{abo-zaid_individual_2013} In particular, when the distribution of the gold standard measurement of the exposure variable varies across studies and the exposure submodel does not account for this, then inadequate predictions will be be made for the unobserved gold standard measurements. We may model the varying prevalence of the gold standard measurement of the exposure $x$ by applying random intercepts to the exposure model, replacing equation \ref{eq:x_z} with:
\begin{equation}
\label{eq:x_j+z}
    \begin{split}
        x_{ij} & \sim \textnormal{Bernoulli}(p_{ij}),\\
        g(p_{ij}) & = \gamma_{00} +  \gamma_{0j} + \gamma_{1} z_{ij}, \\
    \end{split}
\end{equation}

\noindent where $\gamma_{00} \sim N(0, \sigma^2_{\gamma_{00}}), \gamma_{1}  \sim N(0, \sigma^2_{\gamma_{1}}), \gamma_{0j} \sim N(0, \tau^2_{\gamma_{0j}})$. Whereas it is common to assume a Normal prior distribution for regression coefficients and the intercept, \cite{gustafson_measurement_2003} the choice for a prior distribution for the variance parameters is less straightforward. A prior with too heavy tails will give too much prior weight on high variance, whereas a prior with thin tails will put to much prior weight on a low variance. \cite{williams_bayesian_2018} We here consider a half-Normal (that is, truncated at zero) distribution for parameters for heterogeneity between studies, namely $\tau^2_{\gamma_{0j}} \sim \textnormal{half-}N(0, \xi_{\gamma_0})$ but would like to highlight that several alternatives have been proposed, such as the half-Cauchy, half-t and inverse-gamma distributions. \cite{gelman_prior_2006, williams_bayesian_2018, polson_half-cauchy_2012} The exposure model's contribution to the likelihood is then given by:
\begin{equation}
      p \big(\gamma_{00}, \gamma_1, \tau^2_{\gamma_{0j}}\big) 
      \prod_j \prod_i p \big(x_{ij}  |z_{ij}, \gamma_{00}, \gamma_1,  \tau^2_{\gamma_{0j}}\big) 
\end{equation}

\subsubsection{Adjusting for between-study heterogeneity in misclassification}
For various reasons, the extent of error in the measurement of the exposure may vary by study in an IPD-MA. This may be modeled by applying random intercepts in the measurement model, which can be interpreted as that the log-odds sensitivity and 1 - specificity vary by study. The measurement model is then given by:

\begin{equation}
\label{eq:x*_j+x}
    \begin{split}
        x^*_{ij} & \sim \textnormal{Bernoulli}(p^*_{ij}),\\
        g(p^*_{ij})& = (\lambda_{00} + \lambda_{0j}) x_{ij} + (\phi_{00} + \phi_{0j}) (1 - x_{ij}), \\
    \end{split}
\end{equation}

\noindent where $\lambda_{00} \sim N(0, \sigma^2_{\lambda_{00}}), \phi_{00} \sim N(0, \sigma^2_{\phi_{00}}), \lambda_{0j} \sim N(0, \tau^2_{\lambda_{0j}})$, $\phi_{0j} \sim N(0, \tau^2_{\phi_{0j}})$, $\tau^2_{\lambda_{0j}} \sim \textnormal{half-}N(0, \xi_{\lambda_0})$ and $\tau^2_{\phi_{0j}} \sim \textnormal{half-}N(0, \xi_{\phi_0})$

\subsubsection{Adjusting for participant-specific misclassification}
A more complex situation arises when misclassification is related to participant-level covariates. For instance, recall of exposure values may be poorer in the elderly, the answering of questionnaires may be hampered by poor literacy and  measurement instruments might be designed for specific subgroups of participants. Participant-specific misclassification is particularly problematic if the case-mix distributions vary across studies, as estimates of exposure-outcome associations will then be affected differently across studies. For this reason, the presence of such error can be accounted for by incorporating patient-level covariate effects in the measurement model:

\begin{equation}
\label{eq:x*_j+x+z}
    \begin{split}
        x^*_{ij} & \sim \textnormal{Bernoulli}(p^*_{ij}),\\
        g(p^*_{ij})& = (\lambda_{00} + \lambda_{0j} + \lambda_{1} z_{ij}) x_{ij} + (\phi_{00} + \phi_{0j} + \phi_{1} z_{ij}) (1 - x_{ij}), \\
    \end{split}
\end{equation}

\noindent where $\lambda_{00} \sim N(0, \sigma^2_{\lambda_{00}}), \lambda_{1} \sim N(0, \sigma^2_{\lambda_{1}}), \phi_{00} \sim N(0, \sigma^2_{\phi_{00}})$ and $\phi_{1} \sim N(0, \sigma^2_{\phi_{1}})$. The contribution of the measurement model to the likelihood is then given by:
\begin{equation}
    p \big(\lambda_{00}, \lambda_{1}, \phi_{00} , \phi_{1}, \tau^2_{\lambda_{0j}}, \tau^2_{\phi_{0j}} \big) 
    \prod_j \prod_i p \big(x^*_{ij}|x_{ij}, \lambda_{00}, \lambda_{1},  \tau^2_{\lambda_{0j}}, \phi_{00}, \phi_{1}, \tau^2_{\phi_{0j}}\big) 
\end{equation}

\subsubsection{Accounting for between-study heterogeneity in outcome frequency}
Commonly, in data from an IPD-MA and other clustered data sets the frequency of the outcome varies by study. To account for this effect of clustering within studies, it is generally considered vital that random intercepts for the outcome are applied in an IPD-MA. \cite{abo-zaid_individual_2013} We can add these to the outcome model as follows:

\begin{equation}
\label{eq:y_j+x+z}
    \begin{split}
        y_{ij} & \sim \textnormal{Bernoulli}(\pi_{ij}),\\
        g(\pi_{ij}) & = \beta_{00} + \beta_{0j} + \beta_1 z_{ij} + \beta_2 x_{ij}, \\
    \end{split}
\end{equation}

where $\beta_{00} \sim N(0, \sigma^2_{\beta_{00}}), 
\beta_{0j} \sim N(0, \tau^2_{\beta_{0j}})$, 
$\tau^2_{\beta_{0j}} \sim \textnormal{half-}N(0, \xi_{\beta_0})$, and $\beta_{2} \sim N(0, \sigma^2_{\beta_{2}})$.

\subsubsection{Accounting for between-study heterogeneity in exposure-outcome associations}
Further, the strength of the true exposure-outcome association might also vary by study. To model this, one may adopt a random effects model for the outcome, which does not assume there is a single exposure-outcome association.  \cite{dersimonian_meta-analysis_1986} Instead, it assumes there is a distribution of exposure-outcome associations and it estimates the center and variance of that distribution. 

\begin{equation}
\label{eq:y_j+x+xj+z}
    \begin{split}
        y_{ij} & \sim \textnormal{Bernoulli}(\pi_{ij}),\\
        g(\pi_{ij}) & = \beta_{00} + \beta_{0j} + \beta_1 z_{ij} + \beta_{20} x_{ij} + \beta_{2j} x_{ij},\\
    \end{split}
\end{equation}

\noindent where $
        \beta_{20} \sim N(0, \sigma^2_{\beta_{20}}), 
        \tau^2_{\beta_{20}} \sim \textnormal{half-}N(0, \xi_{\beta_{20}}),
        \beta_{2j} \sim N(0, \tau^2_{\beta_{2j}})$, and
        $\tau^2_{\beta_{2j}} \sim \textnormal{half-}N(0, \xi_{\beta_{2j}})$,
        $\beta_{20}$ is the center of the exposure-outcome association distribution and represents the overall association, $\beta_{2j}$ is the study-specific exposure-outcome association and $\tau^2_{\beta_{2j}}$ is the heterogeneity of the exposure-outcome association across studies. The random effects assumption is commonly adopted in meta-analysis where sources of between-study heterogeneity cannot (fully) be explained using participant-specific information but need to be accounted for. It is also considered a rather safe assumption, as a random effects model will estimate the variance of the exposure-outcome association at near zero when that association does not vary in the sample. Conversely, a common effects model will lead to inadequate estimates when the common effects assumption does not hold. Equations \ref{eq:x_j+z} , \ref{eq:x*_j+x+z} and \ref{eq:y_j+x+xj+z} together are illustrated in Figure \ref{fig:model_basic_and_adv}, and the contribution of the outcome model to the likelihood is then given by:
\begin{equation}
      p \big(\beta_{00}, \beta_1, \beta_2, \tau^2_{\beta_{0j}}, \tau^2_{\beta_{2j}} \big) 
      \prod_j \prod_i p \big(y_{ij}  |x_{ij}, z_{ij}, \beta_{00}, \beta_1, \beta_2, \tau^2_{\beta_{0j}}, \tau^2_{\beta_{2j}}\big)
\end{equation}

The models considered here are identifiable only if sufficient information is present in the data. \cite{richardson_conditional_1993, gustafson_measurement_2003} For instance, to estimate equation \ref{eq:x_z} and \ref{eq:x_j+z} requires that the gold standard measurement of the exposure $x_{ij}$ is observed for sufficient individuals.  Strictly speaking, a single (large) study where the gold standard and surrogate measurements have been observed should be sufficient to estimate the participant-level effects, though more studies would be necessary to estimate the study-level effects. For instance, in our motivating example $x_{ij}$ is available for participants in half of the included studies.

Here we have assumed that the outcome $y$ is available for every participant in every study of the IPD-MA. Though, if unavailable, it could be imputed following equation \ref{eq:y_x+z}, \ref{eq:y_j+x+z} or \ref{eq:y_j+x+xj+z}. To ensure congeniality this imputation model must at least contain the exposure and covariates of the outcome model. \cite{meng_multiple-imputation_1994}

\subsubsection{Accounting for differential error}
So far we have assumed the error in the measurement of the exposure is non-differential, that is that conditional on the gold standard measurement of the value of the exposure and on the perfectly measured covariates, the error in the measurement is unrelated to the outcome. In any other case the error is differential. An example of differential error is recall bias in a case-control (or case-referent) study, where individuals may overestimate (or underestimate) their exposure, as a result of a known outcome. The methods we described can be extended to allow for differential misclassification, by replacing equation \ref{eq:x*_j+x+z} with:

\begin{equation}
\label{eq:x*_j+x+z+y}
    \begin{split}
        x^*_{ij} & \sim \textnormal{Bernoulli}(p^*_{ij}),\\
        g(p^*_{ij})& = (\lambda_{00} + \lambda_{0j} + \lambda_{1} z_{ij} + \lambda_{2} y_{ij}) x_{ij} + (\phi_{00} + \phi_{0j} + \phi_{1} z_{ij} + \phi_{2} y_{ij}) (1 - x_{ij}) \\
    \end{split}
\end{equation}

\noindent where $\lambda_{2} \sim N(0, \sigma^2_{\lambda_{2}}), 
\phi_{2}  \sim N(0, \sigma^2_{\phi_{2}}), 
\tau^2_{\lambda_{0j}} \sim \textnormal{half-}N(0, \xi_{\lambda_{0j}})$ and $\tau^2_{\phi_{0j}}    \sim \textnormal{half-}N(0, \xi_{\phi_{0j}})$. 
This model bears much resemblance to (Bayesian) MI. The difference is that in MI no measurement model is specified and the surrogate measurement instead appears on the right hand side of the exposure model. That is, in the MI approach the surrogate is treated as just another variable, whereas in our approach it is treated as a surrogate of the gold standard. Although this model this model accounts for a form of differential error, it still assumes that the influence of covariates is the same for each level of the outcome and that the random intercept across studies is common for the levels of the outcome. Alternatively, it may be considered more likely that the nature of the misclassification differs entirely for participants with and without the outcome. Similar to differential misclassification in a single study, \cite{goldstein_be_2015} this may be accounted for by stratifying the measurement model for the outcome of interest.

\begin{equation}
    \label{eq:x*_y(j+x+z)}
 \begin{split}
        x^*_{ij}\hspace{2mm} \sim& \textnormal{ Bernoulli}(p^*_{ij}),\\
        g(p^*_{ij})= & (\eta_{00}   + \eta_{0j}   + \eta_{1} z_{ij} ) x_{ij} y_{ij} +  (\theta_{00} + \theta_{0j} + \theta_{1} z_{ij} )  (x_{ij} - 1) y_{ij} + \\
              & (\psi_{00}   + \psi_{0j}   + \psi_{1} z_{ij} )   x_{ij} (y_{ij} - 1) + (\omega_{00} + \omega_{0j} + \omega_{1} z_{ij} )  (x_{ij} - 1) (y_{ij} - 1)
\end{split}
\end{equation}

\noindent where $\eta_{00}   \sim N(0, \sigma^2_{\eta_{00}}), 
\eta_{1}    \sim N(0, \sigma^2_{\eta_{1}}), 
\theta_{00} \sim N(0, \sigma^2_{\theta_{00}}), 
\theta_{1}  \sim N(0, \sigma^2_{\theta_{1}}), 
\psi_{00}   \sim N(0, \sigma^2_{\psi_{00}}), 
\psi_{1}    \sim N(0, \sigma^2_{\psi_{1}}), 
\omega_{00} \sim N(0, \sigma^2_{\omega_{00}}), 
\omega_{1}  \sim N(0, \sigma^2_{\omega_{1}}), 
\eta_{0j}   \sim N(0, \tau^2_{\eta_{0j}}), 
\theta_{0j} \sim N(0, \tau^2_{\theta_{0j}}), 
\psi_{0j}   \sim N(0, \tau^2_{\psi_{0j}}), 
\omega_{0j} \sim N(0, \tau^2_{\omega_{0j}}), 
\tau^2_{\eta_{0j}}   \sim \textnormal{half-}N(0, \xi_{\eta_{0j}}),
\tau^2_{\theta_{0j}} \sim \textnormal{half-}N(0, \xi_{\theta_{0j}}),
\tau^2_{\psi_{0j}}   \sim \textnormal{half-}N(0, \xi_{\psi_{0j}}),$ and $
\tau^2_{\omega_{0j}} \sim \textnormal{half-}N(0, \xi_{\omega_{0j}})$.  In case the error is assumed to be restricted to participants with (or without) the outcome, equation \ref{eq:x*_y(j+x+z)} could easily be simplified by letting $x^*_{ij} = x_{ij}$ for these cases.

\section{Motivating example: application of methods to dengue IPD-MA}
\label{sec:example_denv:application}
To illustrate the impact of misclassification on observed exposure-outcome associations in an IPD-MA, we apply several modeling strategies to estimate the muscle pain-dengue association in patients suspected of dengue. Hereto, we generated three scenarios for a dengue IPD-MA  using real data on dengue as described in section \ref{sec:example_denv:intro}. In all scenarios we allowed the true prevalence of muscle pain and the true misclassification rates to vary across studies. In the first scenario we defined the heterogeneity parameters such that all studies have the same (true) prevalence of dengue conditional on the exposure and the covariate and the same (true) exposure-outcome association of muscle pain, conditional on the covariate. In the second scenario we allowed for heterogeneity in the true prevalence of dengue conditional on the exposure and the covariate but not in the true exposure-outcome association, conditional on the covariate. In the third scenario we allowed for the presence of heterogeneity in both the true prevalence of dengue conditional on the exposure and covariate as well as the true exposure-outcome association of muscle pain,  conditional on the covariate.  

We aim to highlight the ability of the methodology we have presented here to restore this association and its uncertainty, while simultaneously accounting for the clustering of participants within studies and allowing for heterogeneity in the muscle pain-dengue association. 

\subsection{Methods}
We apply eleven Bayesian binary logistic modeling strategies to estimate the muscle pain-dengue association and its heterogeneity across studies. First, we model the full data with a mixed effects model as if the gold standard measurement was observed for all participants in all studies. In reality, this would not be possible as the gold standard would not be observed for some participants, but here it serves as a reference for comparison with the models that are restricted to the observed data. Second, we apply a mixed effects model on the subset of the data for which the gold standard measurement of the exposure was observed, that is, we apply a so-called complete case analysis. Third, we apply a naive mixed effects modeling strategy, in which we take the surrogate measurement as a proxy for any participant for whom the gold standard measurement is not observed. Finally, we apply the 8 models described in section \ref{sec:meth;subsec:ipd}. These models range from not accounting for heterogeneity and accounting for the simplest form of misclassification to accounting for heterogeneity in all submodels and for a differing extent and nature of misclassification. Although many more combinations of the submodels exist, for brevity we chose to apply them in the order as outlined, which results in eight full models for accounting for misclassification. We note that some alternative specifications would not be sensible, as the exposure model needs to contain at least the variables that are included in the outcome model. We use prior distributions as described above, except that we applied inverse-gamma distributions for the parameters for heterogeneity across studies.

We estimated all the models with a Gibbs sampler with two independent chains. After 1000 adaptation and 1000 warm up samples, 25000 samples for the estimation of the parameters were performed in each chain. To reduce autocorrelation, we thinned the samples by a factor 5. The presented estimates are based on the remaining $2 \times 5000$ samples. The code for our motivating example is available on Github (\href{https://github.com/VMTdeJong/Misclassification-Dengue}{github.com/VMTdeJong/Misclassification-Dengue}).

\subsection{Results}
In each of the scenarios (see Section \ref{sec:example_denv:intro}), all models yielded positive estimates with 95\% credibility intervals that excluded zero, which in each case may lead to the conclusion that muscle pain is positively associated with dengue. However, we observed considerable differences between the point estimates and estimated 95\% credibility intervals of the different models, especially for the common muscle pain-dengue association.

\subsubsection{Scenario 1: homogeneous conditional baseline prevalence and exposure-outcome associations across studies}
In the first scenario, the estimated association (log-odds ratio) between muscle pain and dengue in the full data was 0.82 (95\% CI: 0.67 : 0.98, Table \ref{tab:dengue:mod_ests_sce1}). The complete case analysis (0.64, 95\% Credibility Interval: 0.41 : 0.87) and especially the naive analysis (0.47, 95\% CI: 0.34 : 0.60) underestimated this association. The misclassification methods were able to restore the muscle pain-dengue association to various degrees. The model comprising equations \ref{eq:x*_j+x+z},    \ref{eq:x_j+z} \& \ref{eq:y_x+z}, which was the correctly specified model, estimated the log odds ratio for the association at 0.72 (95\% CI: 0.54 : 0.90). Surprisingly, the underspecified misclassification models estimated the association with similar or even less error. The overspecified (i.e. models with excess parameters) misclassification errors estimated the association with a larger error, though the errors were still smaller than the naive and complete case analyses.

\input{Ch_me/denv/ex_fits/ex_fit_1}

All models estimated the between-study heterogeneity of the muscle pain-dengue association very well, as the estimates were very similar to the reference estimate of 0.05 (95\% CI: 0.02 : 0.14) in the full data. The exception was the 95\% CI of the complete case analysis, which was wider (0.02 : 0.23) than the 95\% CI for the other models. This is unsurprising as it uses only a subset of the available data.

\subsubsection{Scenario 2: heterogeneous baseline prevalence across studies}
In this second scenario, the estimated association (log-odds ratio) between muscle pain and dengue in the full data was 0.76 (95\% CI: 0.61 : 0.92, Table \ref{tab:dengue:mod_ests_sce2}). Again, the complete case analysis (0.66, 95\% CI: 0.42 : 0.89) and naive analysis (0.56, 95\% CI: 0.42: 0.70) underestimated this association. The misclassification models all estimated the common muscle pain-dengue association with less error than the naive and complete case analysis. The model comprising equations  \ref{eq:x*_j+x+z},    \ref{eq:x_j+z} \& \ref{eq:y_j+x+z} (i.e. the correctly specified model)  estimated the association at 0.74 (95\% CI: 0.58 : 0.91), which was nearly identical to the estimates by the analysis on the full data. Also, all misclassification models had narrower 95\% Credibility Intervals than the complete case analysis.

All considered models estimated the (lack of) between-study heterogeneity in the muscle pain-dengue association adequately. In the analysis on the full data this heterogeneity was estimated at 0.07 (95\% CI: 0.02 : 0.22). Again, the 95\% CI for the complete case analysis was the widest (95\% CI: 0.02 : 0.33).

\input{Ch_me/denv/ex_fits/ex_fit_2}

\subsubsection{Scenario 3: heterogeneous baseline prevalence and exposure effects across studies)}
In this final scenario, the analysis on the full data yielded a muscle pain-dengue association of 0.87 (95\% CI: 0.60 : 1:14), whereas the complete case analysis estimated it at 1.02 (95\% CI: 0.67 : 1.38, Table \ref{tab:dengue:mod_ests_sce3}) This neatly illustrates that the error in the muscle pain-dengue association estimated by complete case analysis is caused by an increased variance rather than bias, as the estimate by the complete case analysis is now increased with respect to the analysis on the full data, whereas in the other scenarios it was underestimated. As expected, the naive analysis underestimated the association yet again, at 0.60 (95\% CI: 0.31 : 0.89).

Three of the misclassification models' point estimates were further away from the point estimate by the full data than the complete case analysis' point estimate, which highlights that applying a misclassification model is not guaranteed to reduce the error in the point estimate. Yet, these were all underspecified models that did not account for the various forms of heterogeneity. The correctly specified model, comprising equations \ref{eq:x*_j+x+z},    \ref{eq:x_j+z} \& \ref{eq:y_j+x+xj+z} estimated the muscle pain-dengue association at 0.79 (95\% CI: 0.48 : 1.11), which was close to the estimate on the full data. The overspecified models yielded similar estimates.

Except for the complete case analysis, all models that estimated the between-study heterogeneity for the muscle pain-dengue association yielded adequate estimates for this variance. The complete case analysis underestimated the amount of between-study heterogeneity, whereas the underspecified misclassification models (wrongly) assumed it to be equal to 0.

\input{Ch_me/denv/ex_fits/ex_fit_3}

\subsection{Summary} 
Overall, the results of this motivating example on the association between muscle pain and dengue highlight the impact of misclassification on an exposure-outcome association. The misclassification models estimated the exposure-outcome association with less error (where the full data is taken as reference) than both the complete-case and naive approaches, with the exception for some models that were underspecified in scenario 3. This suggests that even in these scenarios for relatively small IPD-MAs, the more complex (possibly overspecified) models seem more suitable than the simpler (possibly underspecified) models. 

In general, the models provided adequate estimates of the heterogeneity of the muscle pain-dengue association. The exception was the complete case analysis, which yielded different point estimates due the fact that these estimates were based on different data and which yielded wider credibility intervals due to the fact that these interval estimates were based on less data. In conclusion, the misclassification methods that accounted for heterogeneity in the various submodels gave the best available estimates of the muscle pain-dengue association and its heterogeneity.

\section{Simulation study}
We performed a simulation study to assess the impact of misclassification on estimated exposure-outcome associations and the heterogeneity thereof in an IPD-MA and to assess the validity of our methodology. We aim to highlight the bias that occurs in an exposure-outcome association when misclassification is not accounted for and the ability of the methodology we have presented here to  provide (possibly) unbiased estimates of these associations while propagating the uncertainty induced by misclassification and the various forms of heterogeneity, to facilitate valid inference.

\subsection{Simulation methods}
The data were simulated with a data generating mechanism similar to that in scenario 3 of the motivating example considering the diagnosis of dengue: there was heterogeneity in the distribution of the exposure of interest (muscle pain), in the true prevalence of the exposure conditional on the exposure and covariate, and in the true exposure-outcome association conditional on the covariate. Though, for the purpose of the simulation we used different values for several parameters. The default parameters were (i.e. scenario 1) were 
$n_j = 500,
J = 10, 
J_{gold} = 5, 
k = 1,
\beta_{00} = -0.5, 
\tau_{\beta_{0j}} = 0.25, 
\beta_x = 1, 
\tau_{\beta_x} = 0.15, 
\beta_{z_k} = 0, 
\gamma_{00} = -0.25, 
\tau{\gamma_{0j}} = 0.25, 
\gamma_{z_k} = 1  \textnormal{ for } k \in (1,3),  \textnormal{ and }         \gamma_{z_k} = -1 \textnormal{ for } k \in (2,4), 
\lambda_{00} = 3,     
\tau_{\lambda_{0j}} = 1, 
\lambda_{z_k} = 1  \textnormal{ for } k \in (1,3),  \textnormal{ and } 
\lambda_{z_k} = -1 \textnormal{ for } k \in (2,4), 
\phi_{00} = -3,     
\tau_{\phi_{0j}} = 1, 
\phi_{z_k} = 1  \textnormal{ for } k \in (1,3)  \textnormal{ and } 
\phi_{z_k} = -1 \textnormal{ for } k \in (2,4)$, where $z_k$ is the k-th normally distributed covariate (dropping the $ij$ subscripts for simplicity here). Further, for these covariates $z_k$ the standard deviation of the mean across studies (i.e. heterogeneity) equaled $0.5$, and the covariance of all $z$ within studies equaled $0.25$. In each subsequent scenario, we altered the value for one parameter (Table \ref{sim:tab:par}). We analyzed the estimates for the summary and heterogeneity estimates of the exposure-outcome association ($\beta_x$ and $\tau_{\beta_x}$, respectively) for each model in terms of bias and root mean square error (RMSE), and coverage probability of the 95\% Credibility Interval  relative to the true values. We performed one thousand replications of the simulation in \texttt{R} 3.5.2. \cite{r_core_team_r_2020} In replications where at least one method failed to produce a result, we removed the results for all methods. 

\input{Ch_me/sim/parameters}

In each repetition of the simulation we applied five models on the simulated data, of which three used the IPD and two used aggregated data. First, we applied a model on the complete cases, that is only on the participants for whom the gold standard exposure was observed. Second, we applied a naive model in which the surrogate measurement of the muscle pain was used for participants for whom the gold standard measurement was not available. Third, we applied the misclassification model given by equations \ref{eq:x*_j+x+z}, \ref{eq:x_j+z} \& \ref{eq:y_j+x+xj+z}. Fourth, we applied (frequentist) logistic regression models on the separate data sets for which the gold standard was available, and then performed a Bayesian meta-analysis on the aggregate data (the logistic regression model coefficients and standard errors). In the fifth analysis, we repeated this aggregate data meta-analysis, but now on the data from all studies and used $x^*$ where $x$ was unavailable.

We estimated all the models with a Gibbs sampler with two independent chains using JAGS 4.3.0. After 1000 adaptation and 5000 warm up samples, 10000 samples for the estimation of the parameters were performed in each chain. To reduce autocorrelation, we thinned the samples by a factor 2. The presented estimates are based on the remaining 2 * 5000 samples. The code for our simulation study is available on Github (\href{https://github.com/VMTdeJong/Misclassification-IPDMA}{github.com/VMTdeJong/Misclassification-IPDMA})

\subsection{Simulation results}
\subsubsection{Summary estimate of the exposure-outcome relation}
\input{Ch_me/sim/plot_beta}
As expected, both the IPD and AD analyses that used only data from the studies from which the gold standard $x$ was available, produced (nearly) unbiased results (Figure \ref{sim:fig:results_beta}, top). The exception was scenario 5, where only 100 observations were available per study and the results of all methods were subject to small sample bias. Both of the naive analyses that used $x*$ where $x$ was not available produced biased estimates in all of the scenarios. Though in scenario 4, where in only one study $x$ was not observed, the bias was negligible. In scenario 8, 9 and 10 the covariate-related misclassification was increased, and the naive analyses were the most biased. In scenario 11, where confounding was introduced, the bias was similar to the default scenario. The misclassification model produced (nearly) unbiased results in all scenarios. The exceptions were scenario 5, where all methods were slightly biased due to small samples, and scenario 1 and 2 where we observed nonzero but negligible bias.

Due to the reduced sample size for the methods that used only the gold standard $x$, the variance of the estimates increased, which increased the RMSE (Figure \ref{sim:fig:results_beta}, bottom). As a result, these methods had the largest RMSE in several scenarios where the sample size was small, either due to small samples per study (especially scenario 5, but also 6) or due to few available studies (scenario 2). The RMSE was also high for the naive methods in several scenarios, but especially in scenario's 8, 9 and 10, where misclassification was more frequent. The misclassification had the lowest RMSE in most scenario's, and in others it was tied with other methods. It had especially favourable RMSE when there were few studies with observations of the gold standard (scenario 2), when the sample size was increased (scenario 7) and when misclassification was more frequent (scenario 8, 9 and 10). When in seven out of ten studies the gold standard was observed, the improvement in RMSE was negligible as compared to the other methods (scenario 3). Further, in scenario five, where the sample size was the smallest, the RMSE was similar to that of the naive methods.

The proportion of 95\% Credibility Intervals that covered the true effect size was very high for the methods that used only the data from the studies in which $x$ was observed (Figure \ref{sim:fig:results_beta}, middle). In some scenario's coverage was 100\%, as there was insufficient information to reliably simultaneously estimate the variance of $\beta_x$ and its heterogeneity $\tau_{\beta_x}$. The estimates for the variance were frequently overestimated (not shown), perhaps as a result of the influence from the prior for the variance. The coverage rate for the misclassification model was also too high, though it was closer to nominal. The naive methods had poor coverage in the scenario's where most bias occurred (scenario's 2, 8, 9 and 10), though in some scenario's the coverage was approximately nominal (scenario's 1, 3, 5, 6, 7 and 11).

In conclusion, the misclassification model provided the best available estimates of the exposure-outcome association. Though, in some scenarios there was little or no improvement over other methods. When the number of studies for which $x$ is available is high and the sample size per study is large, analyzing only these studies may be sufficient. However when the sample sizes are small or the number of studies for which the gold standard $x$ is available, applying a misclassification model may reduce the variance of the estimate considerably while retaining (near) unbiasedness.

\subsubsection{Heterogeneity of the exposure-outcome relation}

\input{Ch_me/sim/plot_tau}
All methods produced positively biased estimates of the heterogeneity of the exposure-outcome relation ($\tau_{\beta_x}$) in all scenarios, though the magnitude of the bias varied across methods and scenarios (Figure \ref{sim:fig:results_tau}, top). The methods that used only data from the studies where the gold standard $x$ was measured consistently produced estimates with the greatest bias, though in scenario 4 they were tied with other methods. Surprisingly, the naive methods produced estimates of the heterogeneity of the exposure-outcome relation that in some scenarios were less biased than those from the misclassification model. This was especially the case in scenarios 8, 9 and 10, where we observed the most bias towards zero for the naive methods' summary estimates.  

The methods that only used the data from the studies for which the gold standard was available had the highest RMSE across all scenarios, except when in nine out of ten studies the gold standard was observed (scenario 4), and the difference was small when in seven out of ten studies the gold standard was observed, (scenario 3, Figure \ref{sim:fig:results_tau}, bottom). The naive IPD method and the misclassification model had the lowest RMSE for the heterogeneity estimate. The naive IPD method had slightly lower RMSE when only small samples were available (scenario 5 and 6), whereas the misclassification method had slower RMSE when more covariates were available and covariate-specific misclassification was greater (scenario's 8, 9 and 10). 

For all of the methods, the 95\% Credibility Intervals had below nominal coverage for estimating the heterogeneity of the exposure-outcome relation in all or most of the scenarios (Figure \ref{sim:fig:results_tau}, middle). The coverage was lowest for the naive methods in scenarios 8, 9 and 10, especially for the naive method based on AD.

\section{Discussion}
\label{sec:discuss_misclass}
As measurement error or misclassification may cause bias in estimated exposure-outcome associations, standard errors and between-study heterogeneity in IPD-MA, it is essential to account for this. We have unified methods for misclassification in meta-analysis in a one-stage Bayesian meta-analysis framework. Our methodology allows for the incorporation of covariates on the individual participant level to facilitate valid inference regarding therapeutic and etiologic effects, and added diagnostic and prognostic value. This modeling of the individual participant outcome, exposure and covariate values occurs via three submodels: one for modeling the measurements, one for modeling the (gold standard) exposure and one for modeling the outcome of interest. By doing so, both individual level and study level effects are accounted for in each part of the analysis. This, in turn, may restore the association between the exposure and the outcome.

In our motivating example data sets, the association between muscle pain and dengue was estimated with reduced error by applying the proposed misclassification models with individual participant covariate effects. These models account for the potential between-study heterogeneity in the prevalence of dengue and yielded adequate estimates of between-study heterogeneity of the muscle pain-dengue association.

In our simulations, we considered multiple scenarios where baseline outcome prevalence conditional on covariates and exposure effects were heterogeneous across studies, as well as the exposure-outcome association and the degree of misclassification, and compared the performance of several models. We found that analyses that only used data from studies in which the gold standard was measured, i.e. complete case analyses, produced unbiased summary estimates of the exposure-outcome relation, but did so with considerably increased error unless in at least seven out of ten studies the gold standard was measured. Hence, the feasibility of restricting the analysis to patients with complete data for the (gold standard) exposure will depend on the remaining sample size. If this number is low, the variance of the resulting estimates will be large. In the extreme case, gold standard measurements are entirely unavailable for participants for whom the outcome is available, making this method impossible. In addition, the validity of a complete case analysis may become challenging when patients (or studies) for which only surrogate exposure are available differ with respect to covariates that are not part of the outcome model.  
As expected, our simulations also showed that naively using a possibly misclassified surrogate measurement for the exposure when the gold standard is not available, introduced bias in the estimates for the exposure-outcome association.

This bias could be avoided or mitigated by using our proposed methodology for misclassification models. As the misclassification models use all available data, the resulting standard error is smaller than that of models that use only the observations for which only the gold standard is available. In our simulation, this resulted in estimates that were consistently better than those of the naive and gold standard methods, or that were equally good. However, we did observe bias for estimates of the heterogeneity of the exposure-outcome relation (using all approaches), which is no surprise as estimating this parameter is notoriously difficult in meta-analyses of few studies.  \cite{gelman_prior_2006, langan_comparison_2018} We found that applying a misclassification model on the individual participant data was particularly beneficial when the misclassification was strongly related to covariate values and when the number of studies where the gold standard exposure was measured was low.

In general, overspecification should not induce bias in the estimates, provided that the sample contains enough information to estimate all parameters. Nor should it affect the coverage as the models appropriately account for the uncertainty. However, we stress that if we had applied an underspecified misclassification model, we would expect to have observed (some) bias in the estimates for the exposure-outcome association, as well as less favourable statistical properties in terms of RMSE and coverage. After all, although the misclassification was non-differential given covariates, once those covariates are removed from the model the misclassification may become differential. \cite{carroll_measurement_2006} 

Contrary to our expectations, the fully Bayesian method that naively used the possibly misclassified $x*$ when the gold standard $x$ was not available, was very well able to estimate the heterogeneity in the heterogeneity of the exposure-outcome relation. However, this may have been a result of two biases cancelling each other out, as all methods produced positively biased estimates for the heterogeneity of the exposure-outcome relation due to a limited sample size per study, influence from the priors and a relatively small amount of studies, and the naive methods produced estimates for the summary estimates of the exposure outcome-relation that were biased towards zero. When all estimates in a meta-analysis are drawn towards the null and the SE is kept constant, the heterogeneity estimate is guaranteed to be drawn towards zero.

Another surprise is that the estimation of heterogeneity of the exposure-outcome relation by the naive and misclassification methods was hardly affected by the number of studies for which the gold standard was observed; three out of ten (the lowest in our simulations) was sufficient. This is a sharp contrast with the methods that relied on the gold standard, which needed at least seven studies, or perhaps nine.

\subsection{Limitations and future directions}
Although we recommend the implementation of  misclassification models, an alternative strategy is to implement models that require fewer assumptions and do not depend on Bayesian MCMC sampling methods. Two such methods, RC and MIME, do not specify measurement models and require fewer distributional assumptions, and are therefore described as functional methods \cite{gustafson_measurement_2003} or reclassification methods. \cite{spiegelman_estimation_2000} In contrast, in structural methods such as ours, a  model is specified, which when analyzed with Bayesian methods allows for the appropriate propagation of uncertainty. Though, this requires assumptions on the distribution of the gold standard measurement of the exposure and its surrogate measurement. \cite{gustafson_measurement_2003} However, we focused on the scenario where the exposure is a binary variable that is potentially misclassified, which is common in epidemiology. This binary variable is assumed to follow a Bernoulli distribution, so specification of an exposure model does not add a major assumption \cite{carroll_measurement_2006} aside from congeniality, which is also required for RC and MIME. Although both of these methods have been applied to account for misclassification in single studies, neither has yet been adapted to the heterogeneous setting that is IPD-MA. This would require the specification of multiple heterogeneity parameters. We suggest that further research may focus on integrating these into the IPD-MA framework.

In case the exposure is a continuous variable which has been transformed into a binary variable at a specific cut-off point, alternative assumptions are needed for modeling the distribution of the exposure and its measurement error (see e.g. \cite{gustafson_measurement_2003}). Our method could be further extended in case multiple surrogate exposure measurements are available for some or each participant, by specifying a measurement model for each surrogate measurement.

In the simulation study, we applied only one misclassification model as this simulation was intended as a \textit{proof of concept}, not to assess the relative performance of all the described models in a variety of scenarios.  All of the methods discussed here require covariates that predict the value of the gold standard measurement of the exposure to be fruitful. If the available covariates are not predictive of the missing gold standard exposure or the surrogate exposure, only noise would be added by including individual participant covariate effects in the exposure and measurement models, respectively.

Due to the influence of misclassification on exposure-outcome associations and the presence of between-study heterogeneity, and the increase in parameters that are required to account for this, a larger amount of data are necessary than in an IPD-MA where misclassification is absent. This should be especially the case for the more complex misclassification models. In our simulation study however, three thousand individual participants spread over ten studies was sufficient to obtain unbiased estimates of the exposure-outcome relation and the bias was very small for one thousand participants. In a typical IPD-MA, where the sample size is often much larger, there should be enough information to estimate the more complex misclassification models.

\subsection{Conclusion}
In an IPD-MA, the gold standard measurement of an exposure may be entirely unavailable for all participants in some studies, or unavailable for some participants in all studies, leaving the researcher with only surrogate measurements for these participants. If ignored, this induces bias in the estimated parameters for exposure-outcome associations and other parameters of interest, which must be accounted for. Our Bayesian methodology can be applied to participant level data to reduce the error in the estimate of the exposure-outcome association compared to analyses restricted to participants for whom the gold standard measurement is observed, while appropriately propagating uncertainty for all parameters. This may provide unbiased estimates of the exposure-outcome association and coverage of the true effect by the 95\% CI, provided that the model is specified correctly. 

\section{Highlights}

What is already known?
\begin{itemize}
    \item Misclassification of exposures may bias estimates of exposure-outcome associations.
    \item Methods to adjust for misclassification in single studies or aggregate data meta-analyses are available.
\end{itemize}

What is new?
\begin{itemize}
    \item We present Bayesian methods that allow misclassification of binary exposure variables to depend on study- and participant-level characteristics and that account for residual heterogeneity between studies in an individual participant data meta-analysis (IPDMA).
    \item In an IPDMA with various types of heterogeneity across studies, the proposed methods may produce (nearly) unbiased estimates of the exposure-outcome relation, and the estimates may be more accurate than those from analyses restricted to studies that have only measured the gold standard exposure.
\end{itemize}

Potential impact for other fields.
\begin{itemize}
    \item Misclassification methods may be used to adjust for misclassification of exposures in observational studies in a wide range of fields.
\end{itemize}

\section{Acknowledgements}
\begin{minipage}{.08\textwidth}
\includegraphics[width=\textwidth]{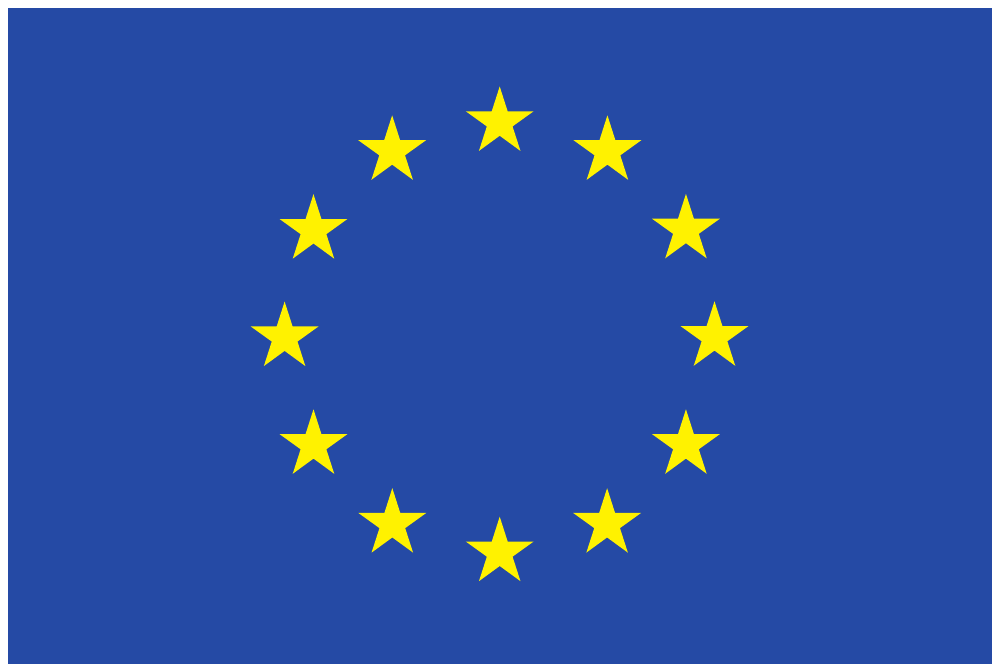}
\end{minipage}
\begin{minipage}{.90\textwidth}
This project has received funding from the European Union's Horizon 2020 research and innovation programme under ReCoDID grant agreement No 825746.
\end{minipage}

We thank the IDAMS consortium for providing aggregate data on the diagnosis of dengue.

\section{Disclaimer}
The views expressed in this paper are the personal views of the authors and may not be understood or quoted as being made on behalf of or reflecting the position of the regulatory agency/agencies or organizations with which the authors are employed/affiliated.

%% file: Ch_me/denv/dengue_plot.tex
\begin{figure}[htbp]
\caption{\enspace Prevalence of dengue and muscle pain measurements in the motivating example}
\label{sim:fig:dengue_bar_chart}
\centering

\includegraphics[width=\linewidth]{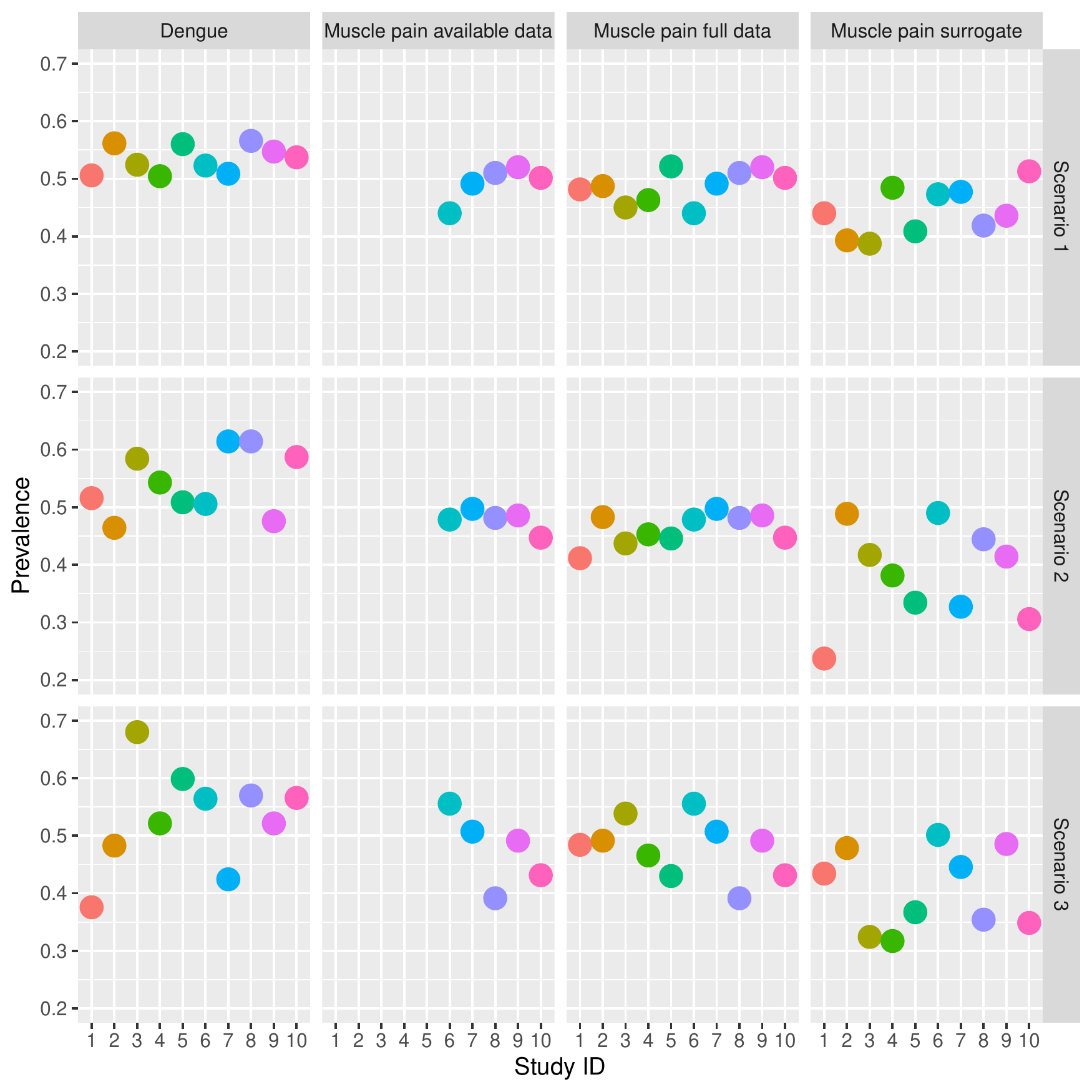}

\footnotesize
Muscle pain was not observed in studies 1 to 5. The values for study 1 to 5 under 'Muscle pain full data' indicate the true values of the prevalence that were not observed.

\end{figure}

%% file: figure_models.tex
\begin{figure}[hptb]
    \caption{\enspace Diagrams of model equations \ref{eq:x*_x}, \ref{eq:x_z} and \ref{eq:y_x+z} (left) and \ref{eq:x_j+z}, \ref{eq:x*_j+x+z} and \ref{eq:y_j+x+xj+z} (right)}
    \label{fig:model_basic_and_adv}
\centering
\includegraphics[width=\linewidth]{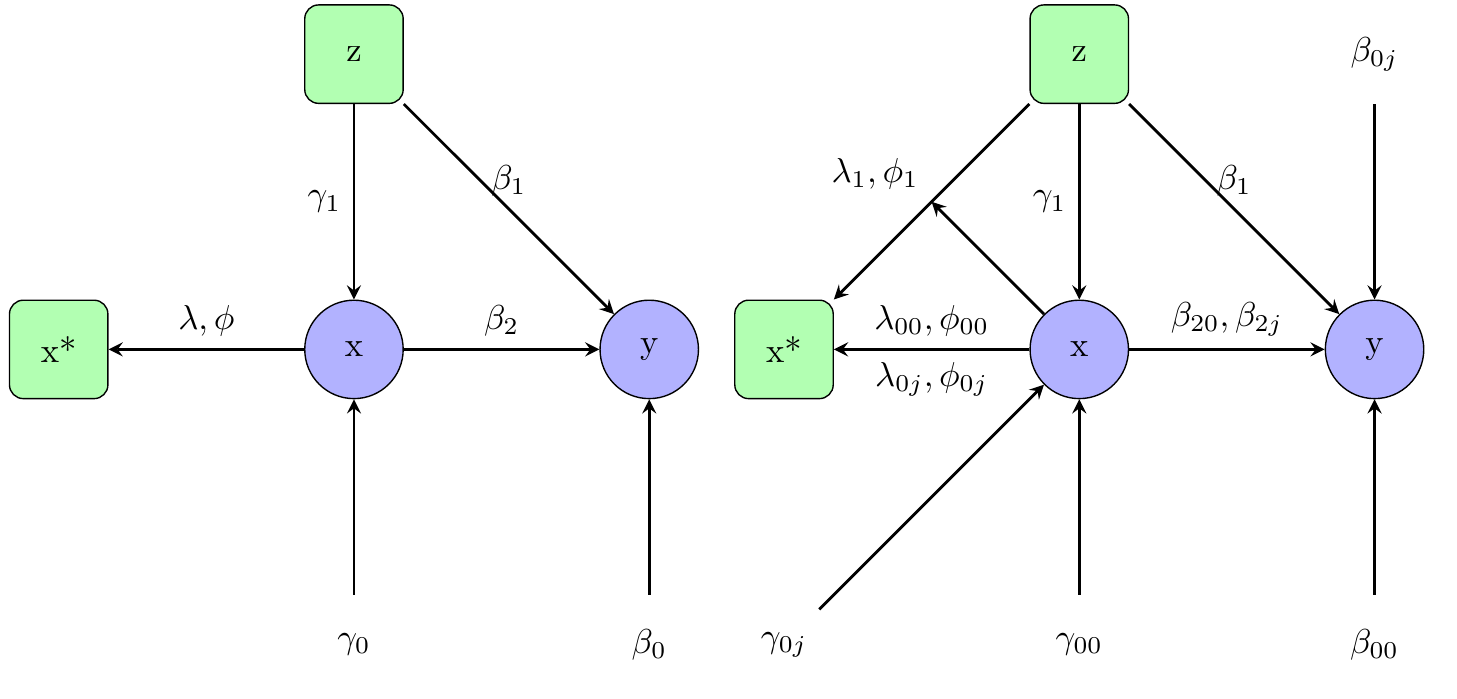}
\footnotesize 
Green squares: fully observed data, blue circles: at least partially observed data, not in boxes: parameters. Variance parameters omitted.
\end{figure}

%% file: Ch_me/denv/ex_fits/ex_fit_1.tex
\begin{center}
\begin{threeparttable}[!h]
\caption{Multivariable log odds ratio and heterogeneity estimates (95\% Credibility Interval) for presence of muscle pain for diagnosing dengue in scenario 1}
\label{tab:dengue:mod_ests_sce1}
\begin{tabular}{lll}
\toprule 
Model              & $\beta_{20} (95\% $CI$)$   & $\tau_{\beta_{2j}} (95\% $CI$)$ \\
\midrule
Full data (reference) & 0.82 (0.67 : 0.98) & 0.05 (0.02 : 0.14) \\
Complete cases        & 0.64 (0.41 : 0.87) & 0.06 (0.02 : 0.23) \\
Naive                 & 0.47 (0.34 : 0.60) & 0.06 (0.02 : 0.16) \\
Equations \ref{eq:x*_x},        \ref{eq:x_z}   \& \ref{eq:y_x+z}       & 0.74 (0.55 : 0.93) &                    \\
Equations \ref{eq:x*_x},        \ref{eq:x_j+z} \& \ref{eq:y_x+z}       & 0.72 (0.53 : 0.91) &                    \\
Equations \ref{eq:x*_j+x},      \ref{eq:x_j+z} \& \ref{eq:y_x+z}       & 0.75 (0.56 : 0.93) &                    \\
Equations \ref{eq:x*_j+x+z},    \ref{eq:x_j+z} \& \ref{eq:y_x+z}       & 0.72 (0.54 : 0.90) &                    \\
Equations \ref{eq:x*_j+x+z},    \ref{eq:x_j+z} \& \ref{eq:y_j+x+z}     & 0.71 (0.54 : 0.90) &                    \\
Equations \ref{eq:x*_j+x+z},    \ref{eq:x_j+z} \& \ref{eq:y_j+x+xj+z}  & 0.71 (0.53 : 0.91) & 0.05 (0.02 : 0.16) \\
Equations \ref{eq:x*_j+x+z+y},  \ref{eq:x_j+z} \& \ref{eq:y_j+x+xj+z}  & 0.70 (0.52 : 0.90) & 0.05 (0.02 : 0.16) \\
Equations \ref{eq:x*_y(j+x+z)}, \ref{eq:x_j+z} \& \ref{eq:y_j+x+xj+z}  & 0.66 (0.45 : 0.88) & 0.05 (0.02 : 0.15) \\
\bottomrule
\end{tabular}
\begin{tablenotes}
    \item The center of the distribution was estimated by the median of the posterior distribution.
    \item Empty cells for $\tau_{\beta_{2j}} (95\% $CI$)$ indicate it is assumed to equal zero in the respective model.
\end{tablenotes}
\end{threeparttable}
\end{center}

%% file: Ch_me/denv/ex_fits/ex_fit_2.tex
\begin{center}
\begin{threeparttable}[!h]
\caption{Multivariable log odds ratio and heterogeneity estimates (95\% Credibility Interval) for presence of muscle pain for diagnosing dengue in scenario 2}
\label{tab:dengue:mod_ests_sce2}
\begin{tabular}{lll}
\toprule 
Model              & $\beta_{20} (95\% $CI$)$   & $\tau_{\beta_{2j}} (95\% $CI$)$ \\
\midrule
Full data (reference) & 0.76 (0.61 : 0.92) & 0.07 (0.02 : 0.22) \\
Complete cases        & 0.66 (0.42 : 0.89) & 0.08 (0.02 : 0.33) \\
Naive                 & 0.56 (0.42 : 0.70) & 0.06 (0.02 : 0.19) \\
Equations \ref{eq:x*_x},        \ref{eq:x_z}   \& \ref{eq:y_x+z}       & 0.75 (0.58 : 0.92) &                    \\
Equations \ref{eq:x*_x},        \ref{eq:x_j+z} \& \ref{eq:y_x+z}       & 0.69 (0.53 : 0.86) &                    \\
Equations \ref{eq:x*_j+x},      \ref{eq:x_j+z} \& \ref{eq:y_x+z}       & 0.76 (0.59 : 0.93) &                    \\
Equations \ref{eq:x*_j+x+z},    \ref{eq:x_j+z} \& \ref{eq:y_x+z}       & 0.73 (0.57 : 0.90) &                    \\
Equations \ref{eq:x*_j+x+z},    \ref{eq:x_j+z} \& \ref{eq:y_j+x+z}     & 0.74 (0.58 : 0.91) &                    \\
Equations \ref{eq:x*_j+x+z},    \ref{eq:x_j+z} \& \ref{eq:y_j+x+xj+z}  & 0.75 (0.57 : 0.94) & 0.09 (0.02 : 0.27) \\
Equations \ref{eq:x*_j+x+z+y},  \ref{eq:x_j+z} \& \ref{eq:y_j+x+xj+z}  & 0.72 (0.54 : 0.91) & 0.08 (0.02 : 0.25) \\
Equations \ref{eq:x*_y(j+x+z)}, \ref{eq:x_j+z} \& \ref{eq:y_j+x+xj+z}  & 0.67 (0.48 : 0.88) & 0.07 (0.02 : 0.23) \\
\bottomrule
\end{tabular}
\begin{tablenotes}
    \item The center of the distribution was estimated by the median of the posterior distribution.
    \item Empty cells for $\tau_{\beta_{2j}} (95\% $CI$)$ indicate it is assumed to equal zero in the respective model.
\end{tablenotes}
\end{threeparttable}
\end{center}

%% file: Ch_me/denv/ex_fits/ex_fit_3.tex
\begin{center}
\begin{threeparttable}[!h]
\caption{Multivariable log odds ratio and heterogeneity estimates (95\% Credibility Interval) for presence of muscle pain for diagnosing dengue in scenario 3}
\label{tab:dengue:mod_ests_sce3}
\begin{tabular}{lll}
\toprule 
Model              & $\beta_{20} (95\% $CI$)$   & $\tau_{\beta_{2j}} (95\% $CI$)$ \\
\midrule
Full data (reference) & 0.87 (0.60 : 1.14) & 0.32 (0.18 : 0.61) \\
Complete cases        & 1.02 (0.67 : 1.38) & 0.23 (0.05 : 0.73) \\
Naive                 & 0.60 (0.31 : 0.89) & 0.37 (0.21 : 0.69) \\
Equations \ref{eq:x*_x},        \ref{eq:x_z}   \& \ref{eq:y_x+z}       & 0.81 (0.63 : 0.99) &                    \\
Equations \ref{eq:x*_x},        \ref{eq:x_j+z} \& \ref{eq:y_x+z}       & 0.58 (0.41 : 0.75) &                    \\
Equations \ref{eq:x*_j+x},      \ref{eq:x_j+z} \& \ref{eq:y_x+z}       & 1.09 (0.92 : 1.28) &                    \\
Equations \ref{eq:x*_j+x+z},    \ref{eq:x_j+z} \& \ref{eq:y_x+z}       & 1.04 (0.88 : 1.22) &                    \\
Equations \ref{eq:x*_j+x+z},    \ref{eq:x_j+z} \& \ref{eq:y_j+x+z}     & 0.97 (0.73 : 1.20) &                    \\
Equations \ref{eq:x*_j+x+z},    \ref{eq:x_j+z} \& \ref{eq:y_j+x+xj+z}  & 0.79 (0.48 : 1.11) & 0.35 (0.19 : 0.67) \\
Equations \ref{eq:x*_j+x+z+y},  \ref{eq:x_j+z} \& \ref{eq:y_j+x+xj+z}   & 0.80 (0.48 : 1.10) & 0.35 (0.18 : 0.68) \\
Equations \ref{eq:x*_y(j+x+z)}, \ref{eq:x_j+z} \& \ref{eq:y_j+x+xj+z}  & 0.82 (0.48 : 1.14) & 0.34 (0.17 : 0.67) \\
\bottomrule
\end{tabular}
\begin{tablenotes}
    \item The center of the distribution was estimated by the median of the posterior distribution.
    \item Empty cells for $\tau_{\beta_{2j}} (95\% $CI$)$ indicate it is assumed to equal zero in the respective model.
\end{tablenotes}
\end{threeparttable}
\end{center}

%% file: Ch_me/sim/parameters.tex
\begin{threeparttable}[!t]
    \centering
    \caption{Simulation Experimental Factors}
    \label{sim:tab:par}
    \begin{tabular}{llll}
    \toprule
    Scenario & Parameter & Value & Interpretation \\
    \midrule
    1     &        &   & Default settings \\
    2     & $J_{gold}$ & 3 & Number of gold standard studies decreased to 3 \\
    3     & $J_{gold}$ & 7 & Number of gold standard studies decreased to 7 \\
    4     & $J_{gold}$ & 9 & Number of gold standard studies decreased to 9 \\
    5     & $n_j$ & 100 & Sample size per study decreased to 100\\
    6     & $n_j$ & 300 & Sample size per study decreased to 300\\
    7     & $n_j$ & 700 & Sample size per study increased to 700 \\
    8     & $K$ & 2 & Number of covariates increased to 2* \\
    9     & $K$ & 3 & Number of covariates increased to 3* \\
    10    & $K$ & 4 & Number of covariates increased to 4* \\
    11    & $\beta_{z_k}$ & 1 & True value for the covariate-outcome relation increased to 1** \\
    \bottomrule
    \end{tabular}
    \begin{tablenotes}
    \item * Covariate-specific misclassification increased accordingly. ** This implies confounding was present.
\end{tablenotes}
\end{threeparttable}

%% file: Ch_me/sim/plot_beta.tex
\begin{figure}[htbp]
\caption{\enspace Bias, Coverage of the 95\% Credibility Interval and Root Mean Square Error (RMSE) for the summary estimate of the exposure-outcome relation}
\label{sim:fig:results_beta}
\centering

\includegraphics[width=\linewidth]{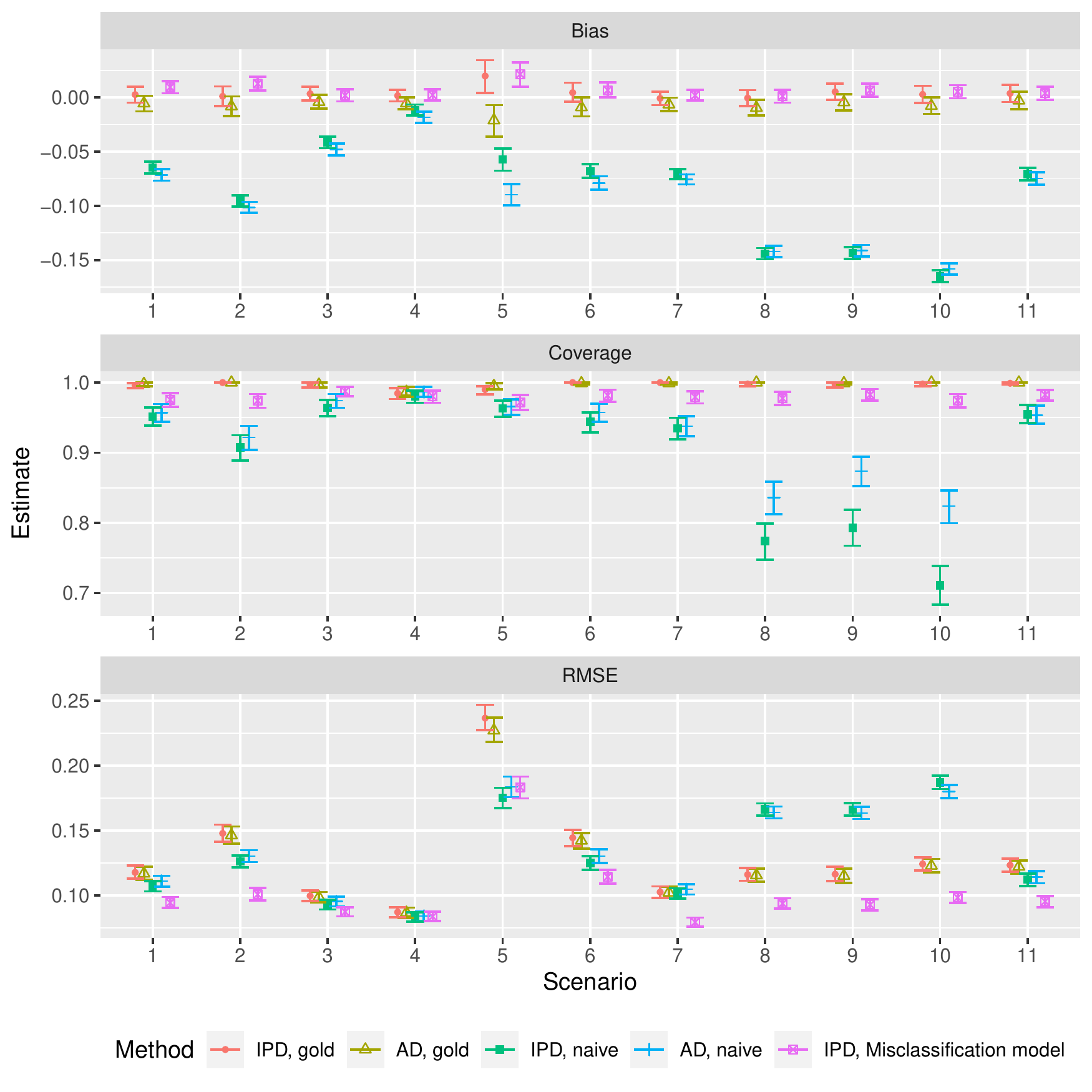}

\footnotesize
IPD: Individual participant data. AD: aggregate data. Gold indicates the model only used studies for which the gold standard $x$ was available. The naive models used $x_*$ for each observation for which $x$ was not available. 
\end{figure}

%% file: Ch_me/sim/plot_tau.tex
\begin{figure}[htbp]
\caption{\enspace Bias, Coverage of the 95\% Credibility Interval and Root Mean Square Error (RMSE) for estimating the heterogeneity of the exposure outcome relation across studies}
\label{sim:fig:results_tau}
\centering

\includegraphics[width=\linewidth]{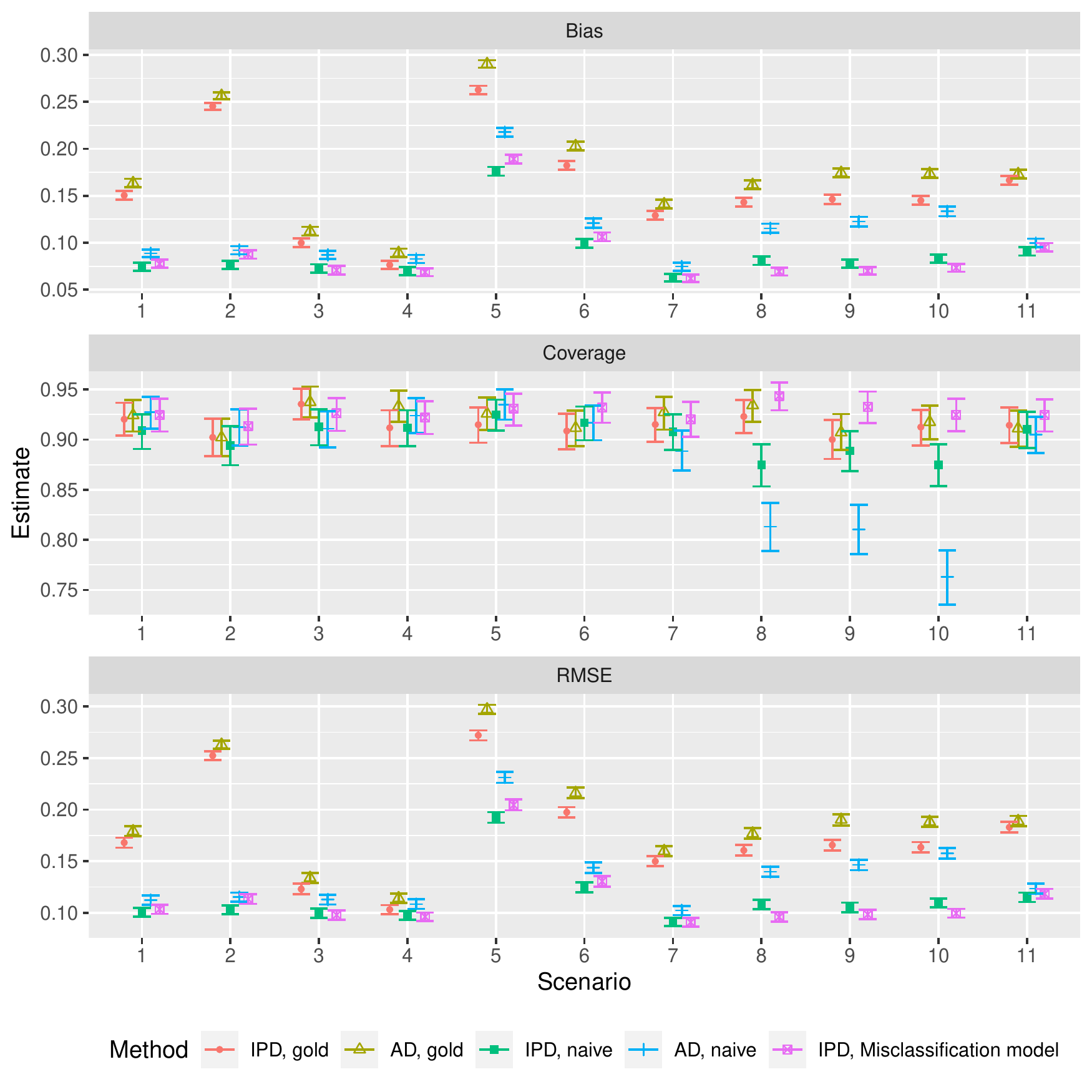}

\footnotesize
IPD: Individual participant data. AD: aggregate data. Gold indicates the model only used studies for which the gold standard $x$ was available. The naive models used $x_*$ for each observation for which $x$ was not available. 
\end{figure}

%% file: Ch_me/denv/appendix_denv_ofi.tex
\section*{Supporting Information: Dengue data}
\label{ch_me:app_dengue_data}
The IDAMS consortium \cite{jaenisch_clinical_2016} provided aggregate data on muscle pain, joint pain and dengue vs other febrile illness (OFI) stratified by three sites (n = 700, 700 and 500), as well as a common association between muscle and joint pain. The IDAMS consortium has collected other clinically important variables, which we do not consider here.

From this data point estimates for the intercept and log odds ratio for the predictor model (equation \ref{eq:x_j+z}) and subsequently the intercept for the outcome model (equation \ref{eq:y_j+x+xj+z}) were estimated with \texttt{optim} in R. \cite{r_core_team_r_2020} As heterogeneity estimates are unreliable in only three sites/studies, we chose suitable values for these for three different scenarios. In the first scenario we set the heterogeneity parameters such that the studies all had identical true incidences of dengue conditional on muscle and joint pain and identical true predictor-outcome (muscle pain-dengue) associations conditional on the covariate, joint pain. In the second scenario we allowed for heterogeneity in the true incidence of dengue conditional on muscle and joint pain but not in the true predictor-outcome association conditional on the covariate, joint pain. In the third scenario we allowed there to be heterogeneity in both the true incidence of dengue conditional on muscle and joint pain as well as the true predictor-outcome association  conditional on the covariate, joint pain. We generated the three scenarios with different simulation seeds, so that unique data sets were generated.

\subsection*{Parameters}
The overall prevalence of joint pain was 0.414. We set the standard deviation for the prevalence of joint pain to 0.1 on the logit scale. The fixed effects for the predictor model were estimated at $\gamma_{00} = -1.70, \gamma_1 = 4.26$. We set $\sigma_{\gamma_{0j}}$ to $0.25$. The fixed effects for the outcome model were $\beta_{00} = -0.26, \beta_1 = -0.06, \beta_2 = 0.78$. We generated data sets for three different scenarios with differing heterogeneity in the outcome model. In scenario 1: $\sigma_{\beta_{0j}} = 0$ and $\sigma_{\beta_{2j}} = 0$. In scenario 2: $\sigma_{\beta_{0j}} = 0.25$ and $\sigma_{\beta_{2j}} = 0$. And in scenario 3: $\sigma_{\beta_{0j}} = 0.25$ and $\sigma_{\beta_{2j}} = 0.15$.

The study specific parameters used to generate the data that was used in the analyses reported in this paper were then generated as follows. We set the number of studies to 10. Study-specific parameters for intercepts and log odds ratios were sampled from normal distributions with their corresponding point and heterogeneity estimates. Prevalences were sampled on the inverse logit scale and then converted to prevalences using the logit function. 

\subsection*{Individual participant data}
Sampling of individual observations was performed using the parameter estimates as follows. We set the sample size to a value similar to the those of the IDAMS consortium: 700 per study, giving a total of 7000 patients. Data for the covariate joint pain were sampled first according to the study-specific prevalences. Then the predictor model (equation \ref{eq:x_j+z}) was applied to sample the muscle pain status. Then the outcome model (equation \ref{eq:y_j+x+xj+z}) was applied to sample dengue status.

The misclassified predictor was generated by equation \ref{eq:x*_j+x+z}, with the following parameter values: $\lambda_{00} = 3, \sigma^2_{\lambda_{0j}} = 1, \lambda_1 = -2, \phi_{00} = -3, \sigma^2_{\phi_{0j}} = 1$ and $\phi_1 = 2.$ The resulting sensitivity and specificity for the sampled true and misclassified muscle pain variables were respectively 0.81 and 0.90 in the full sampled data in scenario 1, 0.78 and 0.96 in scenario 2 and 0.76 and 0.92 in scenario 3. Finally, for the naive and the misclassification methods, for the first five studies the true values for muscle pain were removed, so that only the potentially misclassified values were available for those studies.

\input{Ch_me/denv/table1}

%% file: Ch_me/denv/table1.tex
\begin{center}
\begin{threeparttable}[]
\caption*{Characteristics of dengue data in scenarios 1, 2 \& 3}
\label{tab:dengue:clin_char_123}
\begin{tabular}{llllll}
\toprule
Scenario & Outcome: dengue &     & Absent          & Present         & Total \\
\midrule
1        & Muscle pain$^a$ & Absent  & 1997 (55.6) & 1597 (44.4) & 3594  \\
         &                 & Present & 1267 (37.2) & 2139 (62.8) & 3406  \\
         & Muscle pain$^b$ & Absent  & 968 (54.5)  & 808 (45.5)  & 1776  \\
         &                 & Present & 655 (38.0)  & 1069 (62.0) & 1724  \\
         & Muscle pain$^c$ & Absent  & 2029 (52.0) & 1870 (48.0) & 3899  \\
         &                 & Present & 1235 (39.8) & 1866 (60.2) & 3101  \\
         & Joint pain      & Absent  & 2096 (52.2) & 1920 (47.8) & 4016  \\
         &                 & Present & 1168 (39.1) & 1816 (60.9) & 2984  \vspace{2mm}\\
2        & Muscle pain$^a$ & Absent  & 2024 (53.7) & 1742 (46.3) & 3766  \\
         &                 & Present & 1187 (36.7) & 2047 (63.3) & 3234  \\
         & Muscle pain$^b$ & Absent  & 931 (51.0)  & 896 (49.0)  & 1827  \\
         &                 & Present & 611 (36.5)  & 1062 (63.5) & 1673  \\
         & Muscle pain$^c$ & Absent  & 2195 (50.9) & 2117 (49.1) & 4312  \\
         &                 & Present & 1016 (37.8) & 1672 (62.2) & 2688  \\
         & Joint pain      & Absent  & 2123 (50.7) & 2067 (49.3) & 4190  \\
         &                 & Present & 1088 (38.7) & 1722 (61.3) & 2810  \vspace{2mm}\\
3        & Muscle pain$^a$ & Absent  & 2056 (56.3) & 1593 (43.7) & 3649  \\
         &                 & Present & 1231 (36.7) & 2120 (63.3) & 3351  \\
         & Muscle pain$^b$ & Absent  & 1076 (58.6) & 760 (41.4)  & 1836  \\
         &                 & Present & 572 (34.4)  & 1092 (65.6) & 1664  \\
         & Muscle pain$^c$ & Absent  & 2185 (52.5) & 1975 (47.5) & 4160  \\
         &                 & Present & 1102 (38.8) & 1738 (61.2) & 2840  \\
         & Joint pain      & Absent  & 2184 (53.2) & 1921 (46.8) & 4105  \\
         &                 & Present & 1103 (38.1) & 1792 (61.9) & 2895  \\
\bottomrule         
\end{tabular}
\begin{tablenotes}
    \item Data shown as counts (percentages).
    \item $^a$ As if it were fully observed in all studies.
    \item $^b$ As observed in five studies. Missing in the other five.
    \item $^c$ Potentially misclassified measurement.
\end{tablenotes}
\end{threeparttable}
\end{center}